\documentclass[a4paper,fleqn,usenatbib]{mnras}

\usepackage{newtxtext,newtxmath}

\usepackage[T1]{fontenc}
\usepackage{ae,aecompl}

\usepackage{graphicx}	
\usepackage{amsmath}	
\usepackage{amssymb}	
\usepackage{subfig}

\usepackage{booktabs}
\usepackage{threeparttable}

\title[Multiple populations in M13 with Str\"{o}mgren]{M13 multiple stellar populations seen with the eyes of Str\"{o}mgren photometry}

\author[A. Savino et al.]{
A. Savino,$^{1,2}$\thanks{E-mail: A.savino@rug.nl}
D. Massari,$^{2,3}$
A. Bragaglia,$^{4}$
E. Dalessandro$^{4}$
and E. Tolstoy$^{2}$
\\
$^{1}$Astrophysics Research Institute, Liverpool John Moores University, IC2, Liverpool Science Park, 146 Brownlow Hill, L3 5RF Liverpool, UK\\
$^{2}$Kapteyn Astronomical Institute, University of Groningen, Postbus 800, 9700 AV Groningen, The Netherlands\\
$^{3}$University of Leiden, Leiden Observatory, 2300 RA Leiden, The Netherlands\\
$^{4}$INAF -- Osservatorio Astronomico di Bologna, via Gobetti 93/3, 40129, Bologna, Italy\\
}

\date{Accepted XXX. Received YYY; in original form ZZZ}

\pubyear{2017}

\begin{document}
\label{firstpage}
\pagerange{\pageref{firstpage}--\pageref{lastpage}}
\maketitle

\begin{abstract}
We present a photometric study of M13 multiple stellar populations over a wide field of view, covering approximately 6.5 half-light radii, using archival Isaac Newton Telescope observations to build an accurate multi-band Str\"{o}mgren catalogue. The use of the Str\"{o}mgren index $c_{y}$ permits us to separate the multiple populations of M13 on the basis of their position on the red giant branch. The comparison with medium and high resolution spectroscopic analysis confirms the robustness of our selection criterion. To determine the radial distribution of stars in M13, we complemented our dataset with Hubble Space Telescope observations of the cluster core, to compensate for the effect of incompleteness affecting the most crowded regions. From the analysis of the radial distributions we do not find any significant evidence of spatial segregation. Some residuals may be present in the external regions where we observe only a small number of stars. This finding is compatible with the short dynamical timescale of M13 and represents, to date, one of the few examples of fully spatially mixed multiple populations in a massive globular cluster.
\end{abstract}

\begin{keywords}
 Hertzsprung-Russell and colour-magnitude diagrams -- techniques: photometric -- stars: Population II -- globular clusters: individual: NGC6205
\end{keywords}



\section{Introduction}

Once thought to be the quintessential simple stellar populations \citep{Renzini86}, globular clusters (GCs) have revealed a degree of complexity in their properties that is now challenging our understanding of these celestial objects.

Over the past few decades, a growing body of spectroscopic and photometric evidence has shown the presence of multiple stellar populations in nearly all the Galactic GCs \citep[e.g.][]{Carretta10,Piotto15}, and also in those of nearby galaxies \citep[e.g.][]{Letarte06,Mucciarelli09,Larsen14,Dalessandro16}. Such stellar populations are characterized by marked differences in the chemical abundance of light elemental species. A fraction of stars in GCs present a chemical abundance pattern that is compatible with that of halo stars of the same metallicity. This group of stars is called first, or primordial, population. However, a significant fraction of GC stars show enhancement or depletion of certain light elements (like C, N, O, Na, Mg, Al), in the form of very clear correlations and anticorrelations. These stars, even when belonging to several subgroups, are collectively known as the second, or enriched, population. These light element abundance `anomalies' appear to be linked to an enhancement of the helium content in the chemical mixture from which these stars originated \citep[reviews for photometry and spectroscopy can be found in ][]{Piotto09,Gratton12}.

Many theoretical models have been put forward to try to explain the multiple population phenomenon \citep[e.g., ][]{Ventura01,Decressin07,Demink09,Bastian13,Denissenkov14}. Most of them require the material of second population stars to be first processed by a class of objects called polluters, that are capable of sustaining high temperature CNO cycle reactions. Multiple episodes of star formation are required by some of these scenarios. The precise nature of these polluters is scenario-dependent and it remains a matter of debate. While each model is able to reproduce some of the globular cluster observed properties, each one of them has its own specific limitations \citep[e.g.,][]{Gratton12,Bastian15,Renzini15}. 

A crucial piece of information is thought to reside in the spatial distribution of multiple populations within a cluster. Differences in their radial distribution are expected for most scenarios where multiple star formation events are present. This prediction is complicated by the collisional nature of GCs. Over time, dynamical evolution naturally erases any difference in the initial conditions of the two populations. This effect is faster in the central regions of a cluster, where dynamical timescales  are shorter. However, it has been shown by means of numerical simulations \citep{Vesperini13} that GCs can retain some memory of their primordial configuration, in particular in the external regions.

An ideal dataset to assess this problem would consist of chemical information for a large sample of stars over a wide field of view. While modern multi-object spectrographs have substantially increased the number of spectra, the stellar samples for which we have high resolution elemental abundances are still too small to trace radial distributions in a reliable way.

Photometric studies, on the other hand, have the advantage to efficiently include a much higher number of stars, provided that there is a way to distinguish the different populations. Since the work of \citet{Marino08} and \citet{Yong08} it has become clear that appropriate combination of ultra-violet and optical photometric filters can reveal light element abundance spreads on the RGB. The most powerful tool to perform this kind of analysis is probably represented by the WFC3/UVIS camera, because it offers a filter combination covering features of C, N and O \citep[the last not accessible from the ground -- ][]{Milone12}, on board of the Hubble Space Telescope (hereafter HST). Unfortunately, the size of the HST field of view is small and rarely covers more than one half-light radius for nearby GCs. Fortunately, there are other filter combinations, involving the ultraviolet, which can be efficiently used, in particular narrow and medium-band filters \citep[see, e.g., ][for a thorough study of M5 using a very effective choice of photometric filters]{Lee17}.

In this paper we present the analysis of the multiple populations in the cluster NGC6205 (M13). M13 is a northern GC with a metallicity of [Fe/H] $\simeq -1.55$ dex \citep{Sneden04,Carretta09} that exhibits strong light element variations \citep{Sneden04,Cohen05,Johnson12}. In addition, \citet{Dalessandro13}, by means of synthetic horizontal branch modeling, derived a helium spread of $\Delta Y \simeq 0.05$.

To study multiple populations of M13 we make use of Str\"{o}mgren photometry, probing the cluster out to a distance of 6.5 half-light radii. The power of medium band Str\"{o}mgren filters to identify globular cluster multiple populations has been firstly recognized by \citet{Yong08}. As demonstrated by \citet{Carretta11} and \citet{Massari16}, in this analysis we exploit the efficiency of the $c_{y}$ index in maximizing the effect of carbon and nitrogen abundance variations, to identify the cluster multiple populations and to investigate their spatial distribution. In \S2 we describe the dataset employed for this analysis, in \S3 we present our final photometric catalogue, while \S4  shows the multiple population identification and characterization. We discuss our results and conclusions in \S5.

\section{Data reduction}
\label{Data reduction strategy}
\begin{figure}
	\includegraphics[width=\columnwidth]{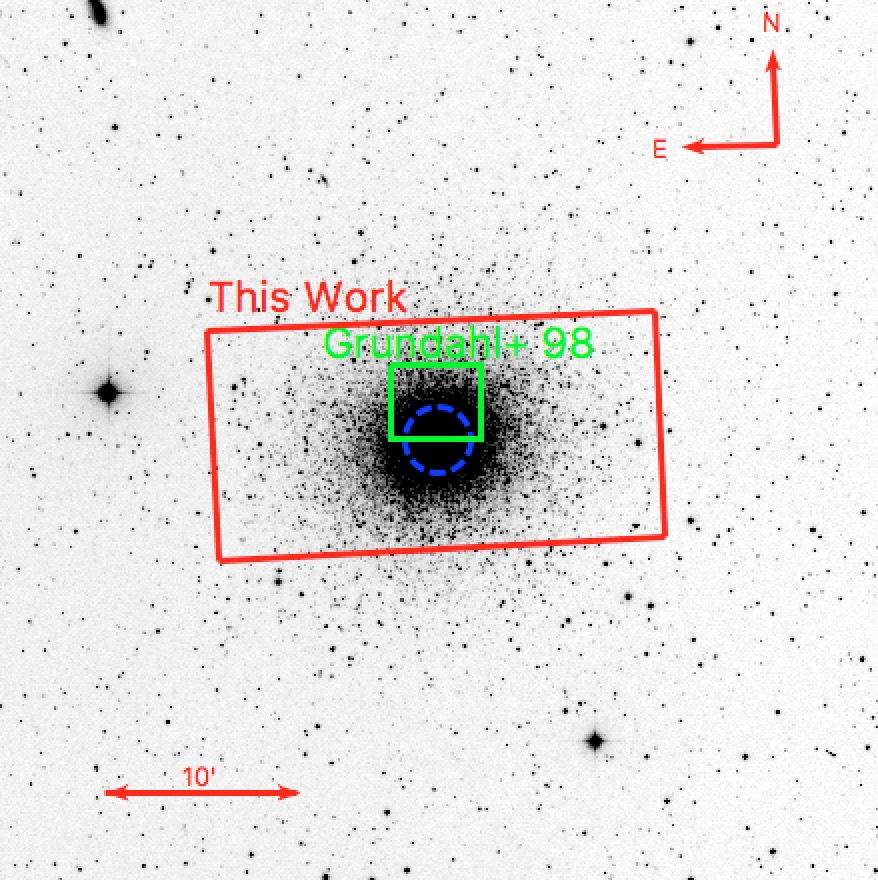}
    \caption{Archival image of M13 from Sloan Digital Sky Survey. The solid red box is the field of view of our dataset for WFC chip four.  The solid green box is the field of view of NOT observations from \citet{Grundahl98}. The dashed blue circle marks M13 half-light radius.}
    \label{fig:FoV}
\end{figure}

Our primary dataset consists of archival images, acquired with the Isaac Newton Telescope - Wide Field Camera (INT-WFC). The INT-WFC has four chips, each covering a field of view of $12.1 \times 23.1$~arcmin, with a pixel scale of 0.33~arcsec. The data for this analysis come only from chip four. Using a single chip prevents any problem related to relative calibration between different detectors, allowing for more precise photometry. This choice is possible as chip four covers most of the extent of M13 ($\sim 6.5$ half-light radii), which is sufficient for our goal. The analysed field of view is shown in Fig.~\ref{fig:FoV}.

\begin{table}
\caption{Exposure times, number of frames and average seeing of our dataset.}
\begin{tabular} {llll}
\toprule
Filter & $t_{exp}$ & $n_{exp}$ & seeing\\
& s & &arcsec\\
 \midrule
  & 60 & 3 & 1.0\\
 \textit{u} & 120 & 4 & 1.2\\
 \hline
  & 45 & 2 & 1.1\\
 \textit{v} & 60 & 3 & 1.2\\
  & 120 & 2 & 1.2\\
 \hline
  & 30 & 2 & 1.2\\
 \textit{b} & 60 & 3 & 1.1\\
  & 80 & 2 & 1.4\\
 \hline
  & 30 & 2 & 1.2\\
 \textit{y} & 60 &3 & 1.1\\
  & 80 & 2 & 1.3\\
 
\bottomrule
\end{tabular}
\label{tab:oblog}
\end{table}
The images have been acquired using the Str\"{o}mgren \textit{uvby} filters, and they come from different observational campaigns, covering different nights from 2004 June 19 to 2013 June 14. A summary of the employed exposures can be found in Table~\ref{tab:oblog}.

The image pre-reduction has been carried out using the {\small IRAF} package\footnote{{\small IRAF} is distributed by the National Optical Astronomy Observatory, which is operated by the Association of Universities for Research 
in Astronomy, Inc., under cooperative agreement with the National Science Foundation. }. For each night, we used ten biases and between five and ten flat-fields per filter. We computed $3\sigma$ clipped median biases and flat-fields using the \textit{zerocombine} and \textit{flatcombine} {\small IRAF} tasks. These images have been used to correct the raw images by means of the \textit{ccdproc} {\small IRAF} task.

The photometric reduction of the scientific images has been carried out using {\small DAOPHOT} and {\small ALLSTAR} \citep{Stetson87} to analyse individual exposures. For each of them, we modelled the point spread function as a Moffat function \citep[with $\beta=2.5;$][]{Moffat69}, using 80 bright, isolated and unsaturated stars. We allowed a linear variation of the model parameters across the field of view. These models were used to build a stellar catalogue for each image. The catalogues were combined using {\small DAOMATCH-DAOMASTER} to estimate mean magnitudes and related uncertainties in each filter. Any zero-point offset caused by different exposure times was corrected by the software, which took into account every star detected in at least two exposures. This choice accounts for pointing differences, maximizing our final field of view.

Among these catalogues, the $u$ filter had the highest number of detections so we used it as a reference to refine the analysis of all our images with {\small ALLFRAME} \citep{Stetson94}. We then used {\small DAOMATCH-DAOMASTER} to create our final multi-band photometric catalogue. Since we want to analyse a combination of different filters, we included only those stars that were detected in all filters.

\begin{figure}
	\includegraphics[width=\columnwidth]{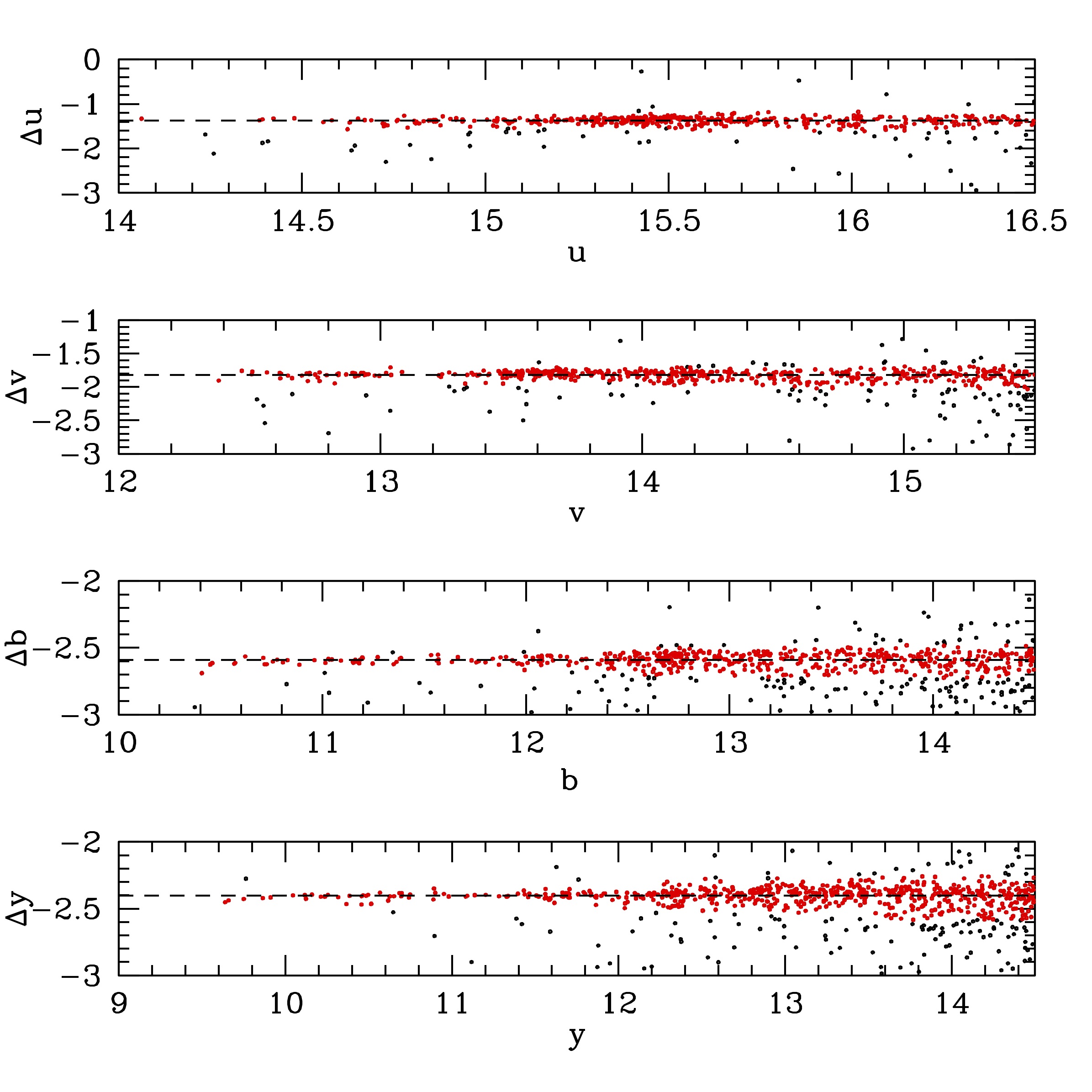}
    \caption{ The magnitude difference as a function of magnitude for the bright stars in common between our catalogue and \citet{Grundahl98}. The four panels show the photometric filters used in this work. The red points are the stars used to calibrate the magnitudes of our dataset. The dashed black lines are the zero-point corrections estimated for our photometry.}
    \label{fig:calib}
\end{figure}

Stellar positions were transformed onto the equatorial reference by using the 2MASS  astrometric system \citep{Skrutskie06}. In order to do so, and for every other cross-match required by our analysis, we used the software {\small CATAPACK}\footnote{www.bo.astro.it/$\sim$paolo/Main/CataPack.html}, developed by P. Montegriffo. For each filter,  the magnitudes have been calibrated onto the Str\"{o}mgren photometric system using the M13 catalogue from \citet{Grundahl98}, made public at http://www.oa- roma.inaf.it/spress/gclusters.html \citep{Calamida07}. The observations had been obtained with the Nordic Optical Telescope (NOT) on a small field of view of about $4 \times 4$~arcmin (see Fig.~\ref{fig:FoV}), with median seeing and pixel size of 0.6 and 0.175~arcsec, respectively. The zero-points were estimated using bright stars in common between the two datasets, adopting a $3\sigma$-clipping procedure to exclude outliers. We verified the absence of possible colour correction terms between the two datasets. Figure~\ref{fig:calib} shows the stars used for the calibration, and the correspondent estimated zero-points for each filter. The complete catalogue will be made available through the CDS.

\begin{figure}
	\includegraphics[width=\columnwidth]{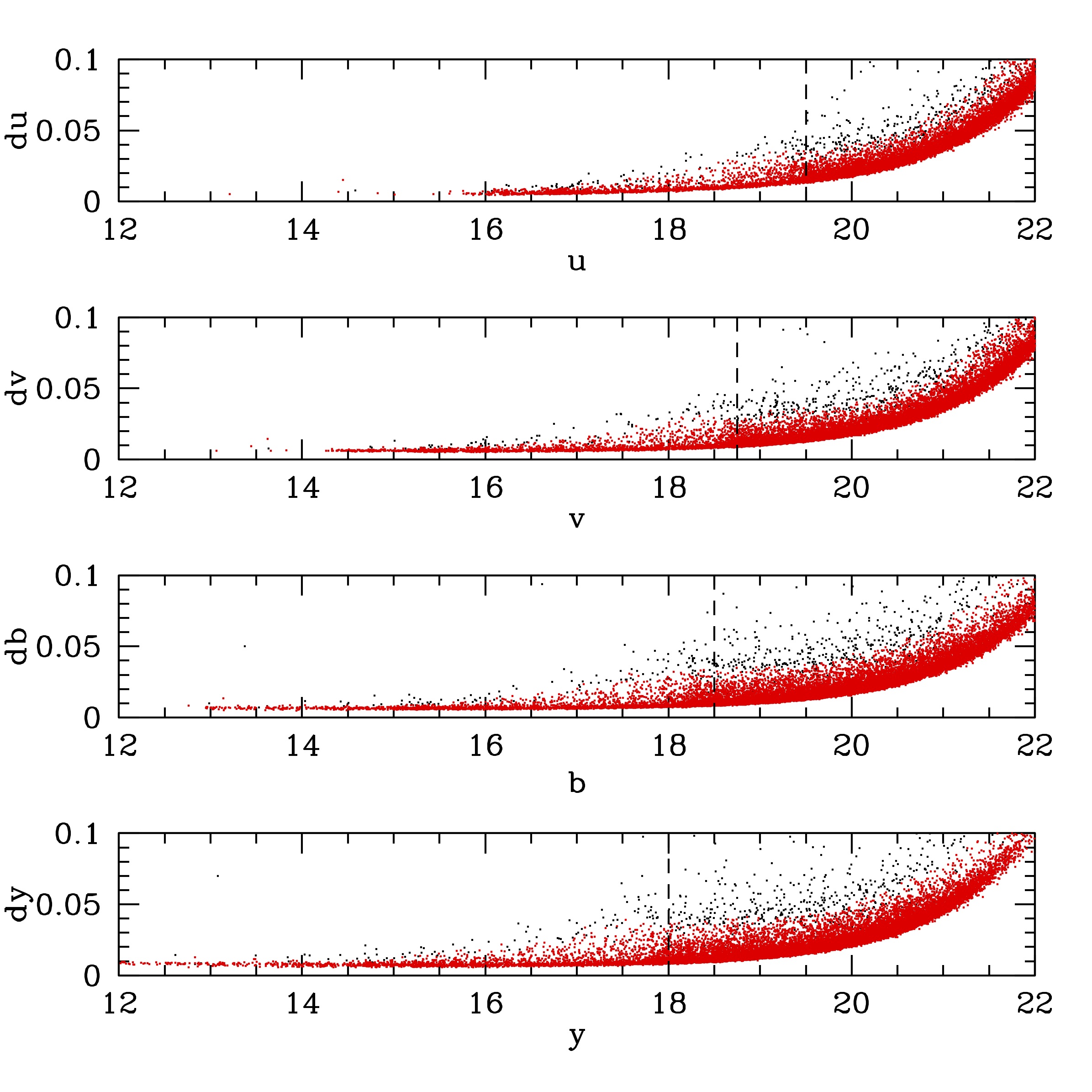}
    \caption{The photometric uncertainty as a function of magnitude for all stars with $\mid\! sharp\!\mid < 0.3$. The red points are the stars selected by our photometric error cut. The dashed vertical lines mark the position of the main sequence turn-off.}
    \label{fig:photerr}
\end{figure}

\section{Colour-magnitude diagram}
\label{CMD}

Our goal is to characterize M13 multiple populations on a photometric basis, and thus precise and reliable magnitudes are essential. For this reason, we applied a quality selection to our catalogue, to exclude stars with bad magnitude estimates, as well as non-stellar objects (background galaxies, cosmic rays, bad pixels, etc.) that were not identified properly during the photometric reduction.

We first applied a selection on the basis of {\small DAOPHOT} sharp parameter. This parameter is defined to be zero for point-like sources. A large positive sharp value typically means that the source is extended. A large negative sharp value means that the object is much smaller than the point spread function width, so it is likely to be an artifact. With that in mind, we excluded all the sources with $\mid\! sharp\!\mid \geq 0.3$.

Figure~\ref{fig:photerr} shows the photometric error distribution as a function of magnitude, after the cut in sharp. We use this distribution to clean our sample further. We include only those stars that, in all four filters, are within three standard deviation from the mean photometric uncertainty at a given magnitude. The stars that do not satisfy this criterion are marked as black points in Fig.~\ref{fig:photerr}.

\begin{figure}
	\includegraphics[width=\columnwidth]{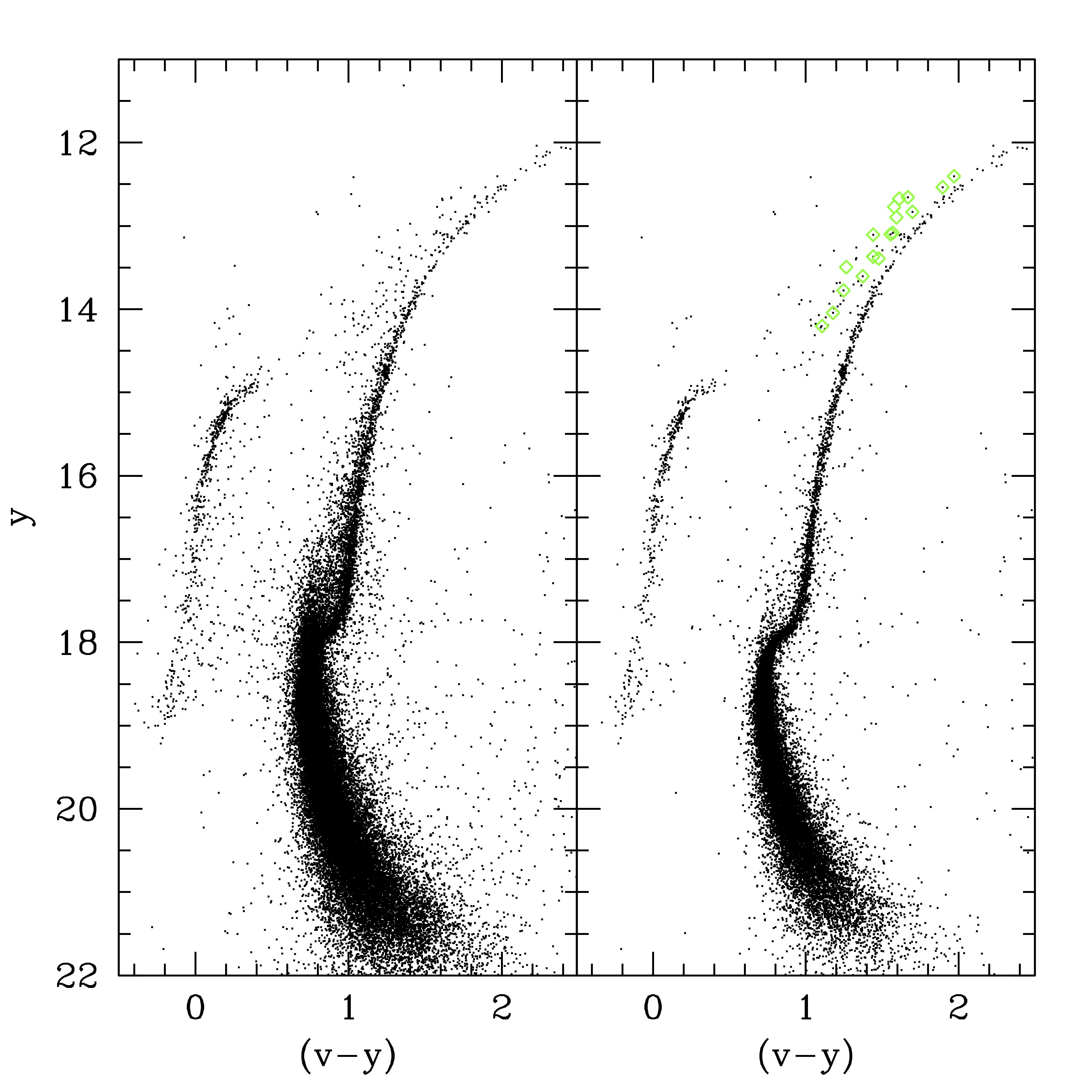}
    \caption{M13 $(v-y)$ vs $y$ CMD. \textit{Left panel:} the entire photometric catalogue. \textit{Right panel:} only the stars selected by our cleaning criteria. Green diamonds are AGB stars with spectroscopic measurements (see \S~\ref{spec} for details).}
    \label{fig:clean}
\end{figure}

\begin{figure}
	\includegraphics[width=\columnwidth]{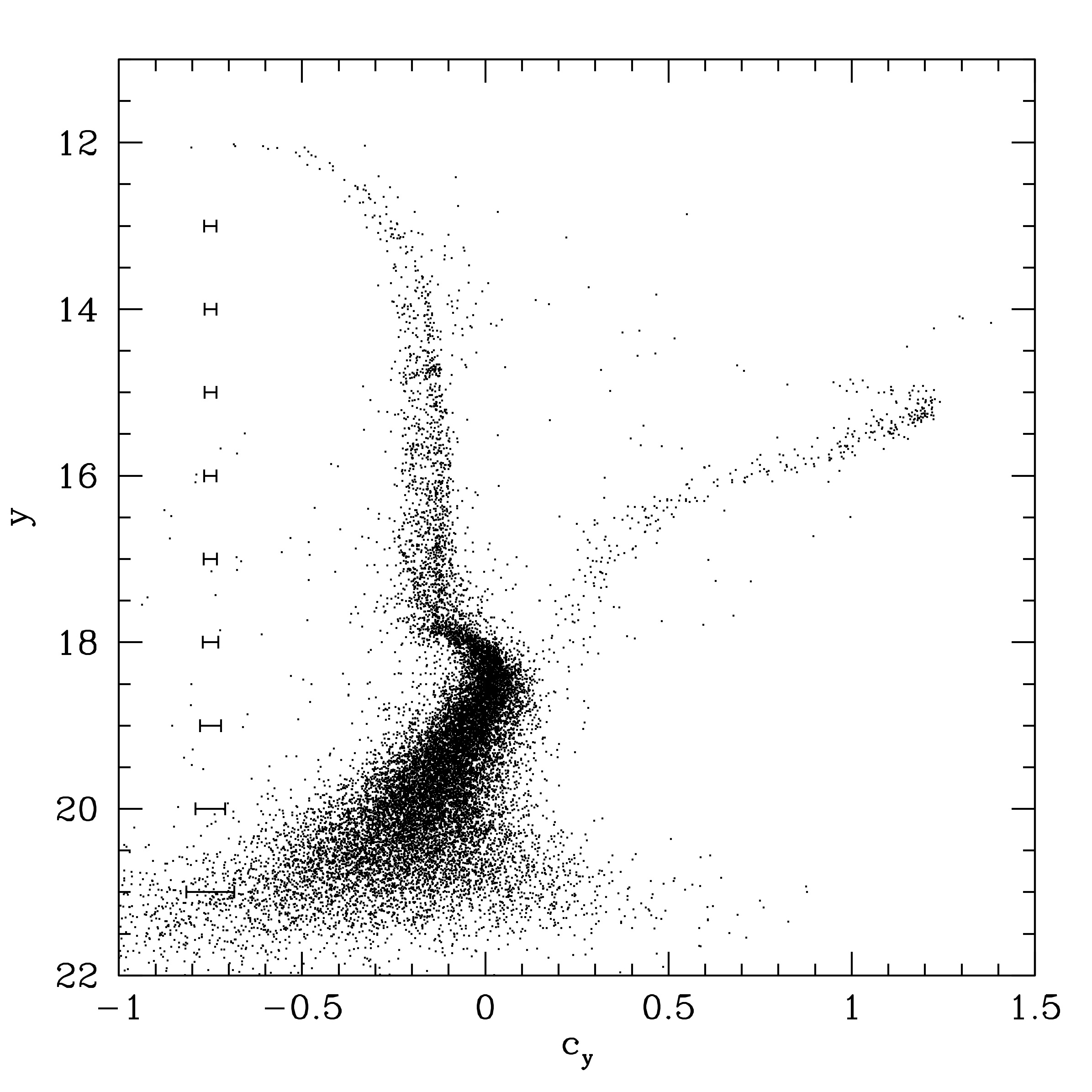}
    \caption{M13 CMD in the $c_{y}$ index versus the $y$ magnitude, extracted from our final catalogue. Average error bars in $c_{y}$ are plotted to permit an easy comparison with the width of the RGB.}
    \label{fig:pseudo}
\end{figure}

Figure~\ref{fig:clean} shows the resulting $(v-y)$ vs $y$ colour magnitude diagram (CMD) before and after the selection. Our cleaning criteria are able to exclude most of stars scattered across the CMD, eliminating most of the potential contaminants. The cleaned sample makes each evolutionary phase more defined, as it can be seen by the thinness of the RGB. 

 Despite the absence of exposures shorter than 30 seconds in our dataset, we are able to recover the magnitude of bright giants up to $y \sim 12$. On the faint end we are able to detect stars $\sim$ 4 magnitudes fainter than the main sequence turn off.

\section{M13 multiple populations}

As it has been demonstrated by numerous studies \citep[e.g.,][]{Yong08,Carretta11,Sbordone11,Roh11,Massari16}, Str\"{o}mgren photometry is a powerful tool to characterize the chemical abundance of GC red giant stars. The small width of Str\"{o}mgren filters, together with their characteristic wavelength, makes them sensitive to the strength of wide molecular features, such as the CN, CH and NH bands \citep[see e.g. fig. 12 in ][]{Carretta11}.

In particular, the $c_{y}$ index, defined by \citet{Yong08} \footnote{$c_y = c_1 - (b - y)$, with $c_1=(u - v) - (v - b) $}, has the advantage of being less sensitive to temperature than the commonly used $c_{1}$ index, and it strongly correlates with the nitrogen abundance. Figure~\ref{fig:pseudo} shows the $c_{y}$ vs $y$ CMD of M13. The RGB clearly splits in two sequences, and the average spread is not compatible with the photometric uncertainties.

\begin{table*}
\centering
\caption{\small Main quantities for the spectroscopic targets in our dataset. Where more than one elemental abundance is present, they are listed in additional rows. The full table is available in the electronic version of the paper.}
\label{tab:spec}
\begin{threeparttable}
\begin{tabular}{rcccccccccc} 
\hline
ID & R.A. & Dec. & $u$ & $v$ & $b$ & $y$  & [O/Fe] & [Na/Fe] & Ref.\tnote{a} & Class\tnote{b}\\
\hline
& (J2000) & (J2000) & mag & mag & mag & mag & dex & dex & & \\
\hline
22037 & 250.4272919 & 36.4489632 & 16.546 & 14.553  &  12.951   & 11.951  &  -0.06  &   0.20  & JP12 & I \\
           &                       &                     &             &              &                &              & -0.13   &   0.27  &   S04               &   \\
22414 & 250.4247742 & 36.4477196 & 16.866 & 14.785  & 13.088    & 12.018  &   0.09  & -0.14   & JP12         & P\\
14695 &  250.3833466& 36.4749908 & 16.435 & 14.541  & 13.000    & 12.041  &   -0.41  &   0.36  & JP12  &  I  \\
           &                       &                     &             &              &                &              &   -0.47   &  0.37&      S04            &   \\
 7188  & 250.3480682 &  36.5047874 & 16.900 & 14.881 & 13.068    & 12.060  &    -1.00   & 0.33 &     JP12          &  E    \\
            &                       &                     &             &              &                &              &   -1.00    & 0.42&      S04            &   \\
            &                       &                     &             &              &                &              &   -1.14    & 0.32  &     CM05          &   \\
17699 & 250.4570923 &  36.4635773 & 16.332 & 14.481 & 12.965    & 12.062   &   -0.56     &0.17   &    JP12              &   I    \\
            &                       &                     &             &              &                &              &   -0.57    & 0.32 &      S04            &   \\
26659 & 250.4459076 &  36.4326668 & 16.378 & 14.524 &  12.959   & 12.077   &   -0.66    & 0.40 &      JP12         &   E    \\
            &                       &                     &             &              &                &              &   -0.70     & 0.46&      S04            &   \\
31813 & 250.3855743 &  36.4118576  & 16.212 & 14.402 &  12.949   & 12.109  &     0.10    &  0.20 &     JP12          &   I    \\
            &                       &                     &             &              &                &              &   -0.10     & 0.39&      S04            &   \\
12557 & 250.4622040 &  36.4818115  & 16.796 & 14.928 &  13.189   & 12.132  &     -1.05    & 0.27&      JP12         &   E    \\
            &                       &                     &             &              &                &              &   -0.78     &0.45 &      S04            &   \\
17099 & 250.3948212 & 36.4665298   &16.251  & 14.439 &  13.041   & 12.153  &    0.00     &0.06             &   JP12            &  I     \\
            &                       &                     &             &              &                &              &   0.00     &0.22  &      S04            &   \\
14609  & 250.4476471 & 36.4745712  &  16.239 & 14.434 & 13.051   & 12.166  &    0.01     &0.20           &    JP12            &  I       \\
            &                       &                     &             &              &                &              &   -0.01     &0.35  &      S04            &   \\

\hline 
\end{tabular}
\begin{tablenotes}
\item[a] Reference for the spectroscopic data: JP12 -- \citet{Johnson12}; S04 -- \citet{Sneden04}; CM05 -- \citet{Cohen05}.
\item[b] Classification according to \citet{Carretta09} criterion: P -- primordial; I -- intermediate; E -- extreme.
\end{tablenotes}
\end{threeparttable}
\end{table*}

\subsection{Spectroscopic confirmation}
\label{spec}

Given the correlation between the $c_{y}$ index and the nitrogen abundance, we expect M13 multiple populations to reside in different regions of the RGB in Fig.~\ref{fig:pseudo}. In particular, we expect N-poor stars to be located on the left side of the $c_{y}$ vs $y$ RGB, while the N-rich giants should occupy the right side. As M13 is a well studied globular cluster, we can confirm this, extending the work done by \citet{Carretta11} on the original NOT data, and trace the $c_{y}$ boundary between different populations, by comparing the results from our catalogue with other independent analyses.

We examine RGB stars for which light element abundances are available. We cross-match our photometric catalogue with three different spectroscopic samples, taken from \citet{Sneden04}, \citet{Cohen05} and \citet{Johnson12}. Across our field of view, we detect 31, 22 and 106 stars, belonging respectively to these datasets.

We use these elemental abundances to separate the different populations, with the criterion introduced in \citet{Carretta09}. We classify our stars into primordial (P), intermediate (I) and extreme (E) populations, based on their [O/Fe] and [Na/Fe] abundances. We call P population those stars with a [Na/Fe] abundance between the minimum measured value [Na/Fe]$_{min}$ (estimated excluding obvious outliers) and [Na/Fe]$_{min} + 0.3$. The I and E populations are defined as the non-P stars with [O/Na] > -0.9 dex and [O/Na] < -0.9, respectively.

 Given the intrinsic differences among the three spectroscopic analyses, such as the spectral resolution and the choice of NLTE corrections, we prefer to apply this criterion separately for each sample, rather than consider all the abundance determinations together. In this way, we avoid the problem of possible zero-point offsets between the different elemental abundance samples. Table~\ref{tab:spec} lists the basic parameters of these spectroscopic datasets.

An additional advantage of treating the samples individually is the presence of stars in common among them, that can be used to check the consistency of the different abundance scales. We have a total of 29 stars with multiple elemental abundance determinations. The catalogues from \citet{Sneden04} and \citet{Johnson12} have 26 stars in common, of which four are also present in \citet{Cohen05}. The common targets between \citet{Cohen05} and \citet{Johnson12} are seven.

 The classification of these stars is in excellent agreement, with only three stars having discordant classifications. Visual inspection reveals that these stars are located near the boundaries, in the parameter space, used to define the different populations. Thus, in principle, these different classifications are compatible with the spectroscopic errors. Given this classification issue we removed these three stars from our sample.

\begin{figure}
	\includegraphics[width=\columnwidth]{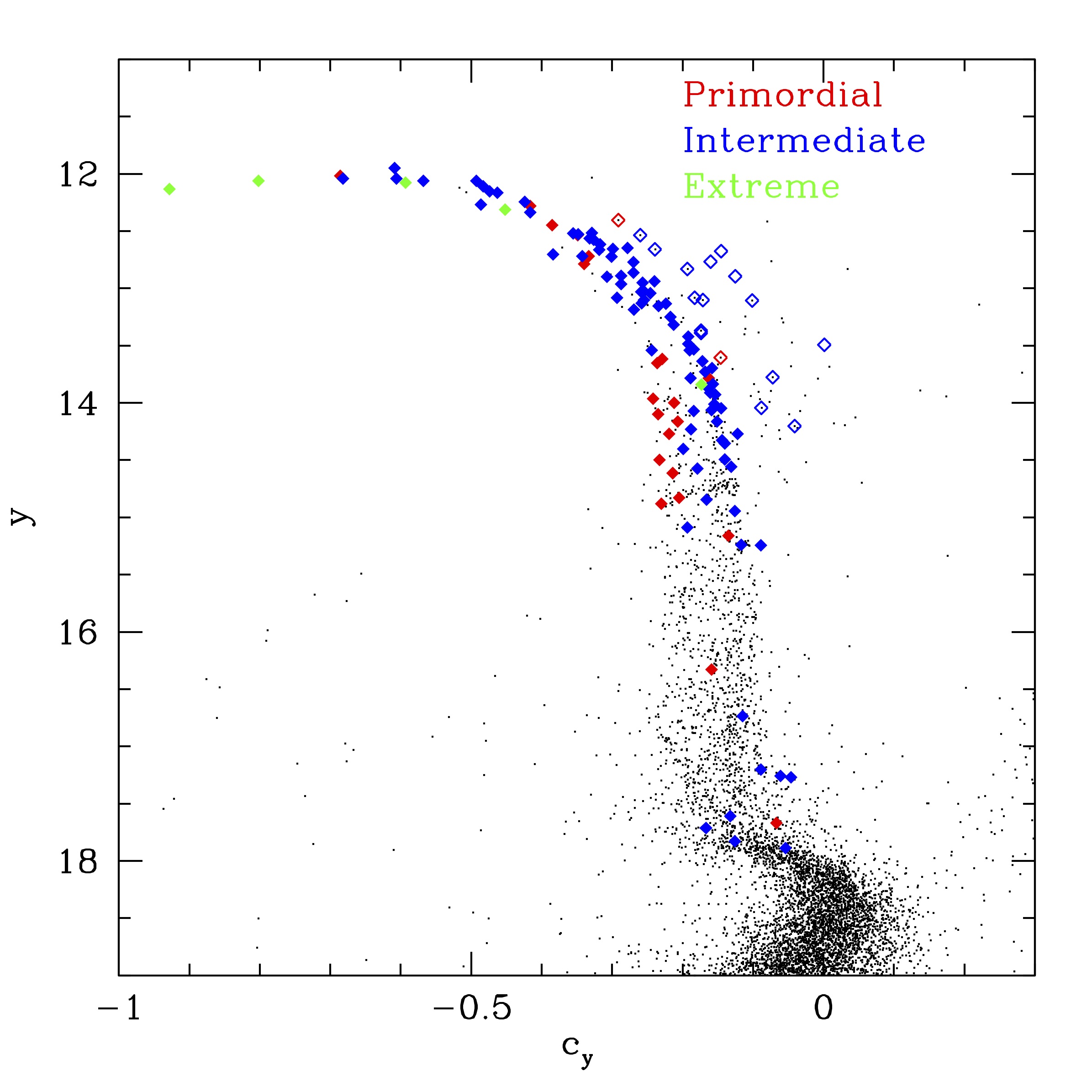}
    \caption{M13 $c_{y}$ vs $y$ CMD (black dots). The diamonds are the stars for which we have spectroscopic data, divided into primordial (red), intermediate (blue) and extreme (green) populations. The empty diamonds are stars that we identify as AGBs.}
    \label{fig:photospec}
\end{figure}

Figure~\ref{fig:photospec} shows the spectroscopic targets, colour coded by population, superimposed to our $c_{y}$ vs $y$ CMD. Unfortunately, most of the elemental abundances are available only for very bright stars, where the pseudo-colour $c_{y}$ is unable to separate the populations. Nonetheless, we reach magnitudes as faint as that of the RGB bump, where a clear separation is seen between the primordial and the enriched populations. It has already been shown \citep{Carretta11}, that the optimal separation in $c_{y}$ is obtained for stars at the level of the RGB bump and below. The result shown in Fig.~\ref{fig:photospec} confirms our ability to use photometry to distinguish between primordial and enriched populations. The mostly clean segregation of the two populations in $c_{y}$allows us to define a photometric criterion to classify M13 stars based on their position on the RGB. As our spectroscopic sample is only representative of stars with $y<15$, the classification of fainter RGB stars unavoidably requires an extrapolation of the trend we observe.

It is interesting that the stars with extreme chemical enrichment are mainly found near the tip of the RGB. This behaviour has already been mentioned by the authors of the three spectroscopic studies \citep[note that the different number of E stars between our work and][ is due to different classification criteria]{Johnson12}. There is only one E star that falls in the magnitude range where the $c_{y}$ spread is significant. As this is not enough to distinguish I and E stars on a photometric basis, in the rest of the analysis we will collectively refer to the I and E stars as the enriched population.

It should be noted that M13 has for long time been the prototype of a GC with very high O depletion, before the analysis of more extreme clusters, in particular NGC2808 \citep{Carretta06}. Such clusters, however, do not show an excess of extreme stars near the RGB tip. In the case of NGC2808, this can be seen in the analysis of \citet{Carretta15}. In spite of tentative explanations \citep[such as the presence of extra mixing in bright stars, suggested by][]{Dantona07}, the unique behaviour of very O-depleted stars in M13 still has to be understood properly. In addition, the presence of kinematic peculiarities in the extreme stars \citep[][]{Cordero17}, makes this population a very interesting subject for further investigation.

As a final remark, we have elemental abundances for a large sample of asymptotic giant branch (AGB) stars (empty symbols in Fig.~\ref{fig:photospec}). The AGB nature of these stars is more easily seen in Fig.~\ref{fig:clean} (green diamonds), where they clearly separated from the RGB. Many of the AGB show [O/Fe] and [Na/Fe] ratios typical of the enriched population. In particular, we classify 15 AGB stars to be part of the enriched population while only two AGB show elemental abundances compatible with the primordial population. While the fraction of enriched stars on the AGB ($88.2^{{+4.7}}_{-23.2}$ per cent) is somewhat higher than that observed on the RGB (see \S~\ref{radial}), the uncertainties coming from the small sample size fully account for this difference. 

This detection is interesting in the light of the recent controversy on the presence of second population stars on the AGB of globular clusters \citep[see, e.g.,][]{Campbell13,Lapenna16,Maclean16,Massari16,Lardo17,Gruyters17}. Stellar evolution theory predicts a fraction of enriched stars on the AGB of GCs, comparable or lower than what observed on the RGB depending on the horizontal branch morphology. However, the presence of these stars is currently debated, with different studies that yield contrasting results. It should be noted, that a scenario where second population stars skip the AGB altogether is hard to reconcile with standard stellar evolution, even when invoking extreme mass loss \citep{Cassisi14}. Our detection in M13 favours enriched stars to ascend the AGB phase, as predicted by stellar population models.

\subsection{Radial profiles}
\label{radial}

The photometric criterion we derived from our comparison with spectroscopy allows us to assign a population membership to RGB stars based solely on their position in the $c_{y}$ vs $y$ CMD (Fig.~\ref{fig:RGBsel}), where we have already applied the refined classification from a comparison with HST data (see below, Fig.~\ref{fig:hstmatch}). In order to avoid foreground contamination we apply the procedure described in \citet{Frank15}, to select genuine RGB stars on the basis of their position in the $(b-y)$ vs $c_1$ and $(b-y)$ vs $m_1$ planes. 

According to this distinction, the fraction of enriched stars in M13 is $74.5^{{+1.9}}_{-2.2}$ per cent. For comparison, the fraction estimated from the spectroscopic sample is $79.8^{{+3.2}}_{-4.7}$ per cent. Before we proceed to the spatial characterization of these populations, some preliminary considerations need to be made regarding the completeness of our stellar catalogue.

\begin{figure}
	\includegraphics[width=\columnwidth]{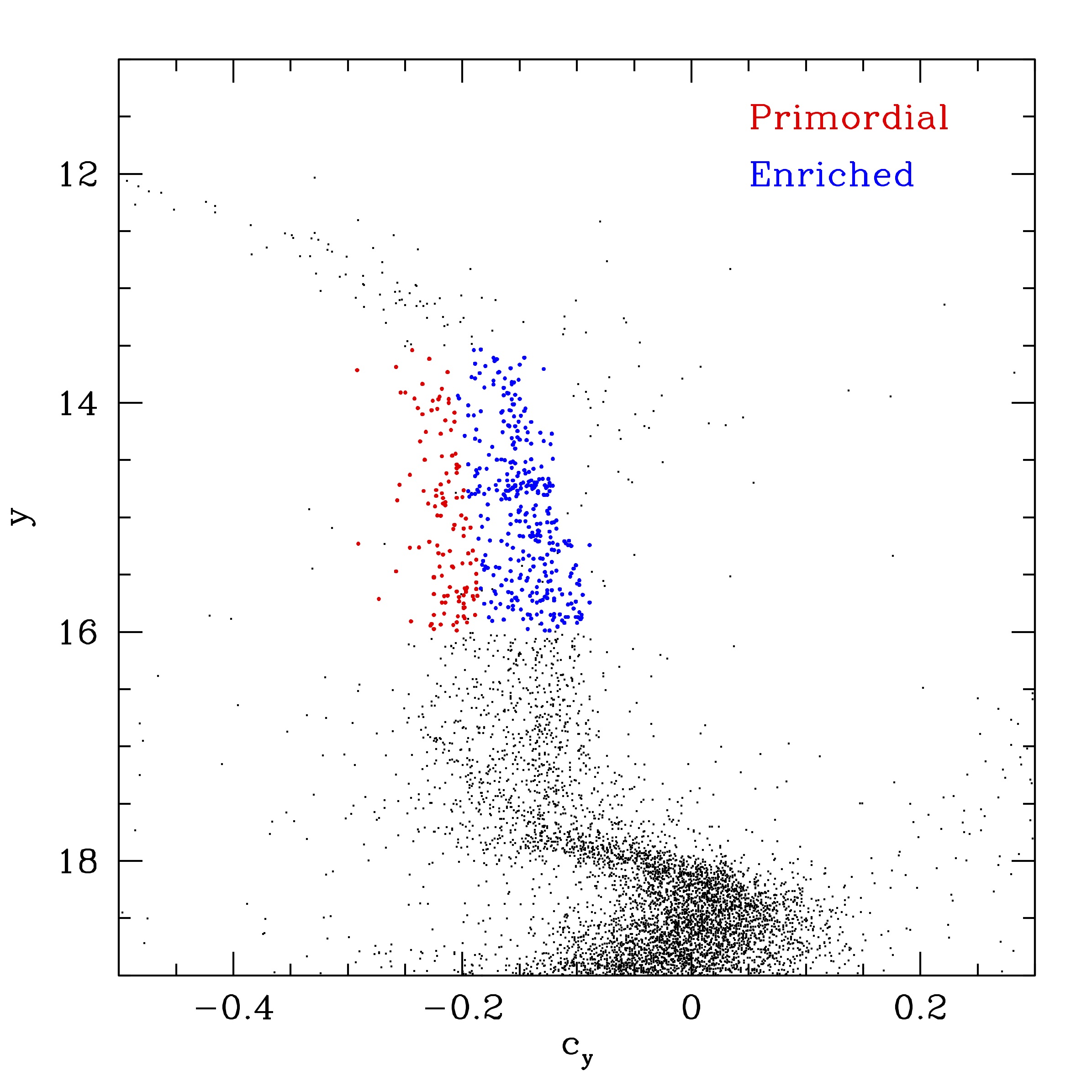}
    \caption{M13 $c_{y}$ vs $y$ CMD (black dots). The red and blue dots represent, respectively, the primordial and enriched population as selected by our photometric criterion. The upper and lower magnitude range for our selection have been obtained by comparison with HST data. See the text for more details.}
    \label{fig:RGBsel}
\end{figure}

For dense stellar systems such as Galactic GCs, the completeness level of post main sequence stars is driven mainly by crowding. Completeness varies with the radius of the globular cluster, with stars preferentially lost in the centre of the cluster rather than in the outskirts. The observed radial distribution is therefore less concentrated than the intrinsic one. For completely mixed populations, this effect is irrelevant, as stars are lost in the same way in the different populations. If one of the stellar populations is more concentrated than the others, it is affected more by incompleteness, and the net effect is a softening in the radial distribution difference.

\begin{figure}
        \subfloat[][]
	{\includegraphics[width=\columnwidth]{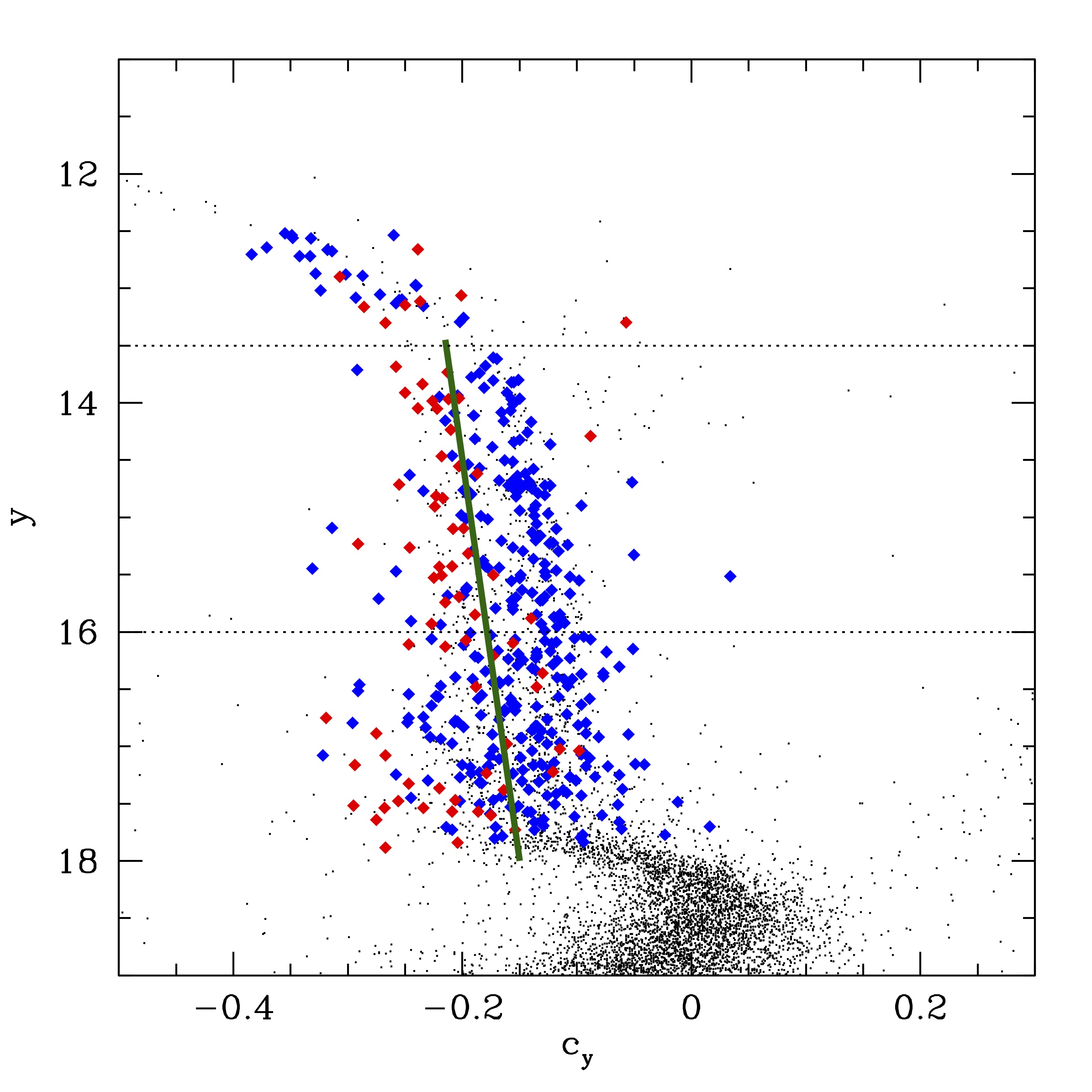}} \quad
	 \subfloat[][]
	{\includegraphics[width=\columnwidth]{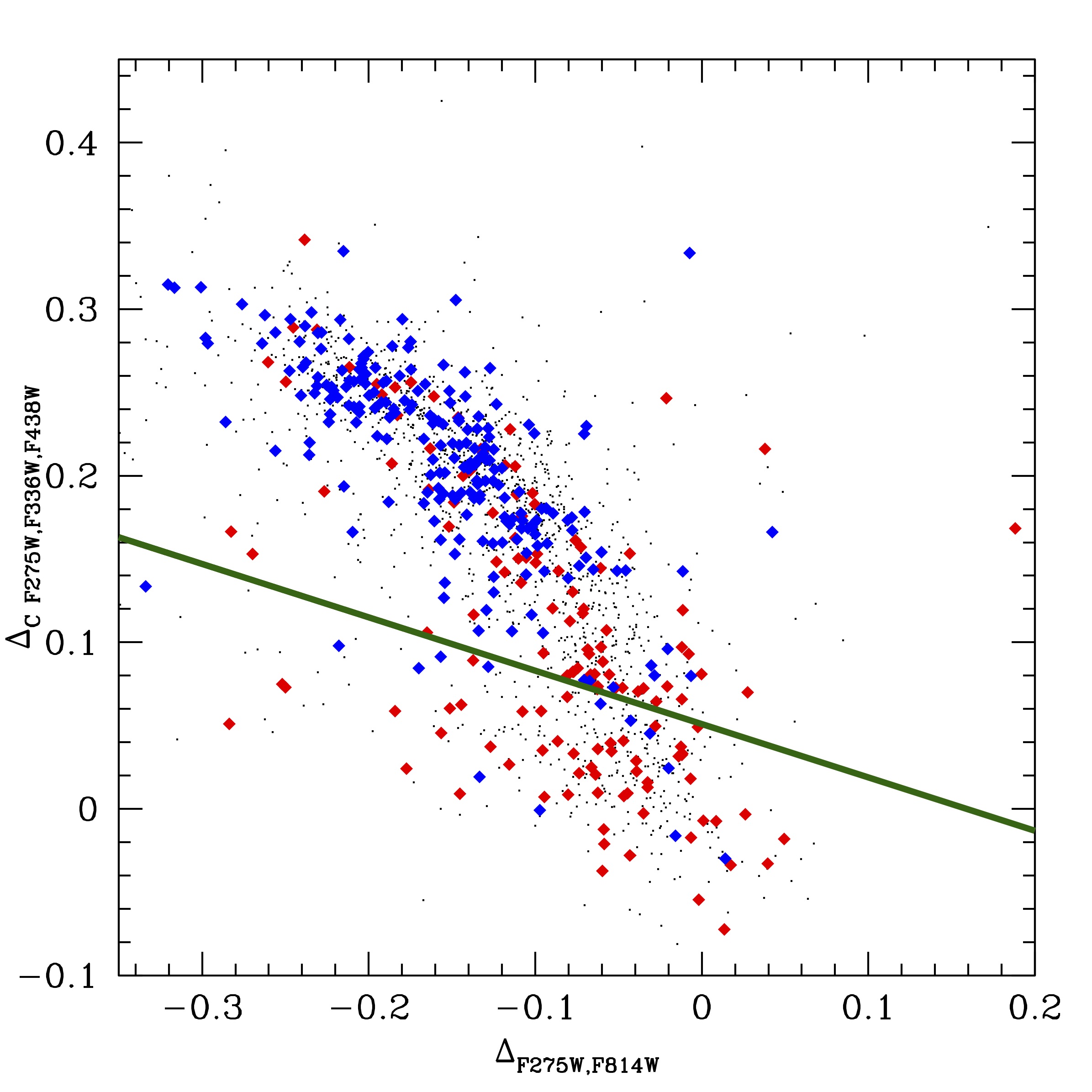}} \quad
    \caption{a) M13 $c_{y}$ vs $y$ CMD from the INT catalogue (black dots). The diamonds are the stars in common with the HST catalogue, divided into primordial (red) and enriched (blue) on the basis of the HST chromosome map criterion. The solid green line shows the photometric criterion defined by equation \ref{eq:phot}. The dotted black lines mark the magnitude range that we use for our analysis. b) Chromosome map built from the HST dataset (black dots). The diamonds are the stars in common with the INT catalogue, divided into primordial (red) and enriched (blue) on the basis of the INT photometric criterion. The solid green line shows the criterion used to separate multiple populations, corresponding to the mean ridge line for the lower sequence, shifted by three times the photometric uncertainty.}
    \label{fig:hstmatch}
\end{figure}

This observational bias can be potentially important for INT observations. While the large field of view makes this telescope a great tool to study the external regions of the cluster, the typical seeing of our images makes it quite sensitive to the crowding of the cluster core.

To assess this issue, we make use of HST observations from the ACS Globular Cluster Survey \citep{Sarajedini07}. We compare our catalogue with HST data in the inner 90~arcsec of the cluster to estimate the difference in the number of detected RGB stars. Even though this comparison only permits us to derive only a relative completeness fraction, the performance of HST in crowded fields is such that nearly all the bright evolved stars are detected, even in the densest region of the cluster.

This instructive comparison allows us to determine that, over the central 90~arcsec of the cluster, the level of completeness of our catalogue is approximately 35 per cent. Such a low fraction of detected stars is mainly the consequence of our very conservative selection criteria. The cuts in sharp and photometric error described in \S~3, while necessary to ensure an accurate characterization of the RGB in terms of $c_{y}$ index, reject a large number of stars with poorly determined magnitudes. This is verified by looking at the completeness level of the catalogue prior to any cleaning selection, and this is nearly 100 per cent.

The low completeness level in the cluster centre makes it very difficult to asses the presence of different radial concentration with the INT observations of this region. For this reason we decided to use our catalogue to study the external regions of M13, while in the centre we make use of more accurate HST observations. As previously demonstrated. \citep[e.g., ][]{Milone12}, HST is a very powerful instrument to study the multiple populations of globular clusters. We thus make use of data from the preliminary data release of the UV Legacy Survey of Galactic Globular Clusters \citep{Piotto15,Soto17}. We follow the procedure described in \citet{Milone17} to distinguish primordial and enriched populations on the basis of M13 `chromosome map' \citep[see fig. 5 of ][]{Milone17}. Chromosome mapping is a technique that exploits the position of stars on the RGB, using different filter combinations, to distinguish between primordial and enriched populations. In particular, the two indices used are sensitive to light element and helium variations respectively. The fraction of enriched stars recovered with this method is $80.4^{{+0.9}}_{-1.0}$ per cent.

Before analysing the multiple population distribution in M13, it is important to asses the consistency of the results obtained with different instruments. Because of the different photometric filters of INT and HST, as well as the different methodologies employed to separate the two populations, we need to verify that the primordial and enriched populations as recovered by HST are consistent with those in our INT data.  We cross-match the two catalogues and asses how the HST-defined multiple populations distribute in the $c_{y}$ vs $y$ CMD, and vice versa (Fig.~\ref{fig:hstmatch}). We found that stars labelled as primordial or enriched on the basis of the HST chromosome map tend to separate in the $c_{y}$ distribution. The agreement with the criterion we defined using spectroscopic data is good, at least for the bright RGB, where INT photometric errors are small enough to permit an accurate classification. We make use of this comparison for a further calibration of our photometric criterion and to avoid extrapolation for stars with $y<15$. In this way the division between primordial and enriched population is given by: 
\begin{equation}
c_{y}=0.014\cdot y + 0.407     
\label{eq:phot}
\end{equation}
which is applied for $13.5<y<16$ and $-0.3<c_{y}<0.07$. The magnitude range is the one for which the HST and INT multiple populations are in good agreement. We note that there are stars for which the two classification criteria don't match. In the colour and magnitude range specified above, the fraction of these stars is around 15\%.

\begin{figure}
	\includegraphics[width=\columnwidth]{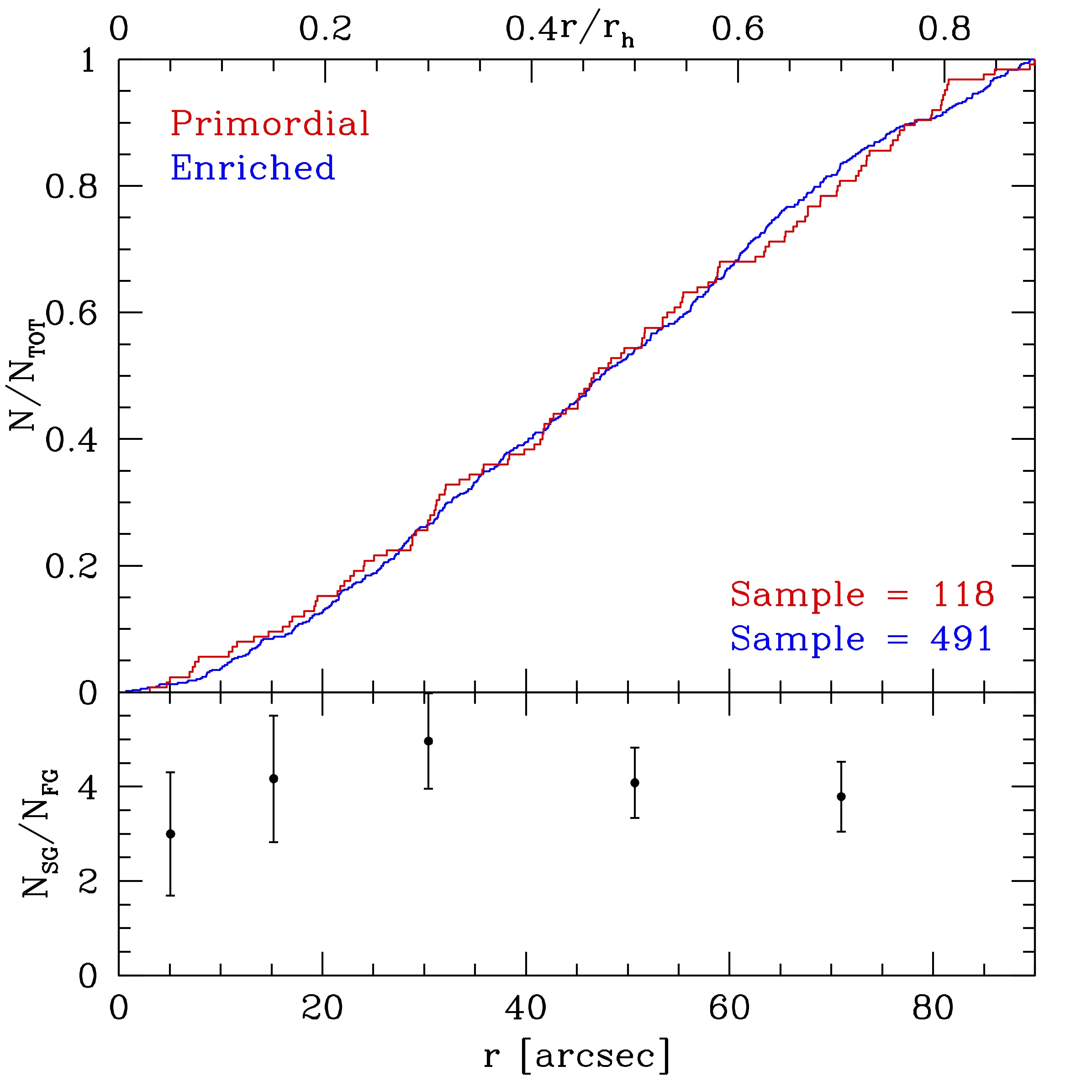}
    \caption{Upper panel: the cumulative distributions for the primordial (red) and enriched (blue) populations in the central regions of M13, obtained using HST data from the UV Legacy Survey of Galactic Globular Clusters. Lower panel:  the number ratio of enriched stars to primordial stars in different radial bins.}
    \label{fig:hstrad}
\end{figure}

\begin{figure}
	\includegraphics[width=\columnwidth]{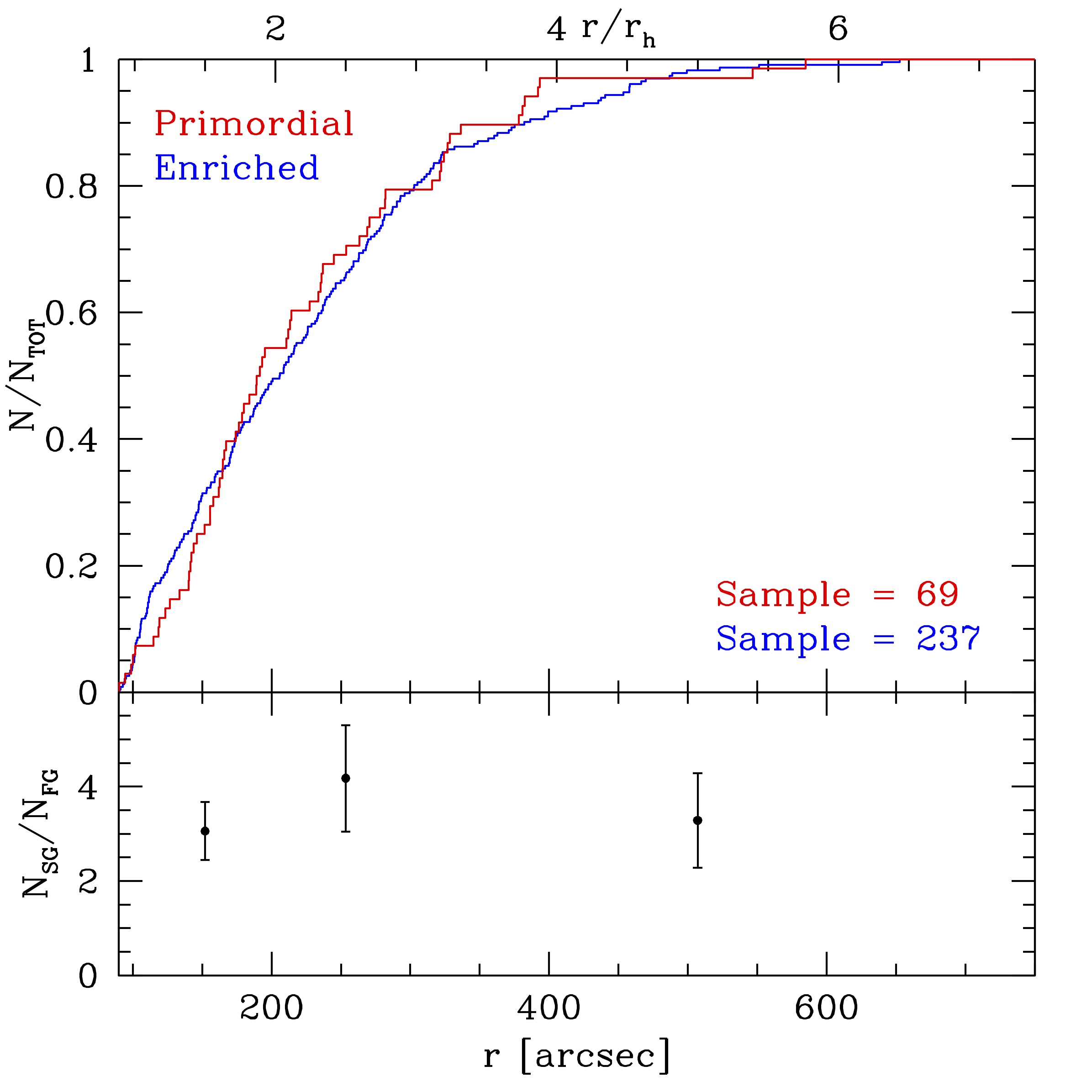}
    \caption{As for Fig.~\ref{fig:hstrad} but for the outer regions of M13. These distributions have been obtained using our INT catalogue.}
    \label{fig:intrad}
\end{figure}

 The radial distribution of the two populations that we find with HST is shown in Fig.~\ref{fig:hstrad}. The two radial profiles are almost indistinguishable, suggesting a complete spatial mixing of the two populations. A Kolgomorov-Smirnov (KS) test gives us a probability of 94.9 per cent that the two distributions are drawn from the same parent population. As mentioned before, the field of view of HST allows us to cover only the inner 90~arcsec of the cluster, which correspond to slightly less than one half-light radius. The complete mixing of the two populations is thus unsurprising, as the dynamical timescales in this region are very short.
 
 The occurrence of complete mixing in the cluster centre simplifies the study of outer spatial distributions. By definition, the shape of a cumulative distribution is influenced by the global properties of the sample. Making a separate cumulative distribution for the outer regions could in principle have led to a biased comparison, as the central properties were not taken into account. However, the high level of mixing in M13 core reassures us that any difference in the global spatial distribution of M13 multiple populations would be driven by the stars in the outer regions. In this context, a comparison of the outer regions alone is relatively safe.
 
 As it can be seen from Fig.~\ref{fig:intrad}, the trend observed in the centre seems to continue also in the outskirts of cluster. The KS probability that the primordial and enriched radial distributions are representations of the same parent population is 65.2 per cent. While this test does not allow us to conclusively exclude the presence of some radial gradients between the multiple populations of M13,  it is fair to conclude that they should be only minor and limited to the most external regions where we only have a small sample of stars.

\section{Discussion and Conclusions}
\label{conclusion}

In this work, we analysed a photometric dataset in Str\"{o}mgren $ubvy$ filters to characterize the multiple stellar populations of the Galactic GC NGC6205 (M13). The use of the $c_{y}$ photometric index revealed the presence of a spread on the RGB, consistent with the chemical inhomogeneity within the cluster.

The comparison with several spectroscopic studies, confirmed that the $c_{y}$ index is a very efficient tracer of the light element abundance in RGB stars, and allows us to distinguish enriched and primordial populations on the basis of their position on the CMD.

Our analysis concluded that around 80 per cent of M13 giant stars belong to the enriched population. This high fraction is in line with the trend of enriched population fraction versus cluster absolute magnitude found by \citet{Milone17}.

To study the radial distribution of M13 multiple populations we complemented our dataset with HST observations from the UV Legacy Survey of Galactic Globular Clusters. The analysis of the radial profiles over a wide field of view revealed no significant evidence for spatial dishomogeneities between the primordial and the enriched population, neither in the inner nor the outer regions of the cluster. 

This result is at odds with what was found by \citet{Lardo11}, who detected a radial gradient in the multiple population distribution of many clusters, including M13. The causes for these discordant results are difficult to asses. However, we note that Str\"{o}mgren filters have a significantly smaller bandwidth and are more sensitive to light element abundances compared to the Sloan passbands used in \citet{Lardo11}. The photometric accuracy of our dataset is also better. Finally, we stress that the classification criteria of both studies have been defined empirically.  
 
 To date, only a few GCs have been observed to have fully spatially mixed multiple populations \citep{Dalessandro14,Nardiello15}. In this light, such a finding for M13 stars is very interesting, especially given that M13 is the most massive cluster in which this phenomenon has been observed.
 
 An interesting comparison can be drawn with NGC5272 (M3). These two clusters are very similar in terms of mass, age and metallicity, and are a classical horizontal branch second parameter problem pair \citep[e.g. ][]{Dalessandro13}. In contrast to M13, M3 multiple populations have been found to be mixed only out to $\sim0.6$ half-light radii, while at larger distance from the cluster centre the enriched population is significantly more concentrated \citep{Massari16}.
  
We suggest that this difference can be understood when considering the dynamical properties of the two clusters. M13's relaxation time at the half-light radius is of the order of 2 Gyrs, which is about three times shorter than that of M3 \citep[][2010 version]{Harris96} and similar to that of the completely mixed NGC6362 \citep{Dalessandro14}.  As the dynamical timescale increases in the less dense regions of a stellar cluster, M3 has achieved a full spatial mixing only in its central part, while the more efficient M13 has also managed to become mixed in its outskirts. Following this argument, we expect that completely mixed stellar populations should be observed in clusters with similar, or shorter, relaxation times (e.g., M71, M30 or NGC 6218).

\section*{Acknowledgements}
The Str\"{o}mgren photometry we employed in this analysis has been obtained from the Isaac Newton Group Archive, which is maintained as part of the CASU Astronomical Data Centre at the Institute of Astronomy, Cambridge. Some of the data presented in this paper were obtained from the Mikulski Archive for Space Telescopes (MAST). STScI is operated by the Association of Universities for Research in Astronomy, Inc., under NASA contract NAS5-26555. Support for MAST for non-HST data is provided by the NASA Office of Space Science via grant NNX09AF08G and by other grants and contracts. This research made use of the SIMBAD data base, operated at CDS, 
Strasbourg, France, and of NASA's Astrophysical Data System. ET thanks INAF-OABO for hospitality.




\bibliographystyle{mnras}
\bibliography{./Bibliography} 

\bsp	
\label{lastpage}
\end{document}